\newif\ifwordcount
\wordcountfalse

\documentclass[%
 reprint,
nofootinbib,superscriptaddress,titlepage,
 amsmath,amssymb,
 prl,
]{revtex4-1}

\usepackage{color}

\usepackage{graphicx}
\usepackage{dcolumn}
\usepackage{bm}

\usepackage{hyperref}
\usepackage{graphicx}
\usepackage{caption}
\usepackage{subcaption}
\usepackage{ragged2e}
\usepackage{multirow}
\usepackage{hhline}
\DeclareCaptionJustification{justified}{\justifying}
\begin{document}

\title{Towards Neutrino Mass from Cosmology without Optical Depth Information}

\author{Byeonghee Yu}\affiliation{Berkeley Center for Cosmological Physics, Department of Physics,
University of California, Berkeley, CA 94720, USA}

\author{Robert Z Knight}
\affiliation{Physics Department, University of California, Davis, CA 95616, USA}

\author{Blake D.~Sherwin}
\affiliation{Department of Applied Mathematics and Theoretical Physics, University of Cambridge, Wilberforce Road, Cambridge CB3 OWA, UK}
\affiliation{Kavli Institute for Cosmology Cambridge, Madingley Road, Cambridge CB3 0HA, UK
}

\author{Simone Ferraro}
\affiliation{Berkeley Center for Cosmological Physics, Department of Physics, University of California, Berkeley, CA 94720, USA}
\affiliation{Miller Institute for Basic Research in Science, University of California, Berkeley, CA 94720, USA}

\author{Lloyd Knox}
\affiliation{Physics Department, University of California, Davis, CA 95616, USA}

\author{Marcel Schmittfull}
\affiliation{Institute for Advanced Study, Einstein Drive, Princeton, NJ 08540, USA}

\begin{abstract}
With low redshift probes reaching unprecedented precision, uncertainty of the CMB optical depth is expected to be the limiting factor for future cosmological neutrino mass constraints. In this paper, we discuss to what extent combinations of CMB lensing and galaxy surveys measurements at low redshifts $z\sim 0.5-5$ will be able to make competitive neutrino mass measurements without relying on any optical depth constraints. We find that the combination of LSST galaxies and CMB-S4 lensing should be able to achieve constraints on the neutrino mass sum of 25meV without optical depth information, an independent measurement that is competitive with or slightly better than the constraint of 30meV possible with CMB-S4 and present-day optical depth measurements. These constraints originate both in structure growth probed by cross-correlation tomography over a wide redshift range as well as, most importantly, the shape of the galaxy power spectrum measured over a large volume. We caution that possible complications such as higher-order biasing and systematic errors in the analysis of high redshift galaxy clustering are only briefly discussed and may be non-negligible. Nevertheless, our results show that new kinds of high-precision neutrino mass measurements at and beyond the present-day optical depth limit may be possible.\end{abstract}
\ifwordcount
\else
\maketitle
\fi

\section{\label{sec:intro}Introduction}

An important goal in both particle physics and cosmology is to understand the physics underlying the neutrino mass \cite{Abazajian:2013oma}. The fact that neutrinos have a non-zero mass has been known since the discovery of neutrino oscillations; however, the absolute scale of this mass is uncertain, with oscillation experiments only giving a lower bound of $\approx 60$ meV for the normal hierarchy and $\approx 100$ meV for the inverted hierarchy. A measurement of the neutrino mass would not just reveal a new energy scale, it would set targets for terrestrial double beta decay experiments (thus potentially contributing to a determination of whether neutrinos are Dirac or Majorana particles) and might even give insight into the mass ordering. Perhaps the most exciting possibility is that the combination of cosmological and laboratory measurements reveals inconsistencies requiring new physics. A cosmological neutrino mass measurement would significantly contribute to efforts to understand physics in the neutrino sector.

The neutrino mass can be probed precisely in cosmology because properties of the cosmic neutrino background affect the growth of cosmic structure and the expansion history of the universe \citep{Lesgourgues:2006nd, Gerbino:2018jee}. A primary effect targeted by future experiments is the suppression of growth of small scale structure caused by a nonzero neutrino mass. The rest mass of the neutrinos, as they become non-relativistic, increases the neutrino contribution to the total mean energy density beyond what it would be in the massless case, thereby increasing the expansion rate and thus suppressing growth. A secondary effect is the scale dependence of this suppression: 
above the free-streaming scale the neutrinos act just like cold dark matter and therefore contribute to gravitational instability, with the net effect of canceling out the suppressive effect of the increased expansion rate \cite{Bond:1983hb, Ma:1996za, Hu:1997vi, Kaplinghat:2003bh, Zhen:2015yba}. This results in a fairly broad ``step''-like feature in the matter power spectrum, where the size of the step is time dependent and grows approximately linearly with every $e$-fold of expansion\footnote{In linear theory, it can be shown that the size of the ``step'' feature in the power spectrum grows by $\frac65 f_\nu$ per $e-$fold of expansion, where $f_\nu = \Omega_{\nu} / \Omega_m$ is the fraction of mass in neutrinos.}.  

To measure neutrino mass using this time-dependent suppression, the amplitude of structure at low redshift (probed by gravitational lensing, clusters, or redshift-space distortions) is typically compared with the initial, high redshift amplitude probed by the CMB. In particular, the small-scale suppression is about 4\% between the redshift of recombination and today for the minimal mass of 60 meV. 

With the design of increasingly powerful CMB surveys, such as CMB Stage-4 experiment (CMB-S4, \cite{Abazajian:2016yjj}) and Simons Observatory (SO, \cite{Ade:2018sbj}), it has become clear that the limiting factor for upcoming neutrino mass constraints will not be the precision of the measurement of CMB lensing or other low redshift probes, but instead the precision of the high redshift amplitude of structure at the CMB redshift $z\approx 1100$ \citep{Allison:2015qca}.
This high-$z$ amplitude $A_s$, in turn, is limited by how well we know the optical depth $\tau$ to the CMB, because the combination $A_s e^{-2\tau}$ (describing the amplitude of the CMB power spectrum) is what is measured by CMB surveys \cite{Allison:2015qca, Mishra-Sharma:2018ykh}. Since it is unclear whether substantially improved $\tau$ constraints will be forthcoming, it is well motivated to seek methods by which the neutrino mass can be probed without relying on a knowledge of the CMB optical depth.

In this paper we examine to what extent the combination of CMB lensing from future experiments with galaxy surveys such as LSST can be used to obtain competitive neutrino mass constraints \emph{without optical depth information}. We further consider what future surveys are required to improve on optical depth limited neutrino mass constraints.

Our investigation is motivated by two effects that may allow neutrino mass constraints without optical depth information. First, the time dependence of the neutrino mass suppression of structure growth can not only be seen by comparing the amplitude of fluctuations of the CMB and today; it can also be seen at low redshift alone using cross-correlations to probe the growth of structure over a sufficiently long redshift lever arm, given sufficiently precise measurements. Extremely high precision constraints on the amplitude of structure as a function of redshift were indeed forecast by \cite{Schmittfull:2017ffw}, and we build on these results in our analysis (for similar recent forecasts also see, for example, \cite{Banerjee:2016suz,Modi:2017wds}). 
Second, there are other physical effects, such as the step feature in the shape of the matter power spectrum described previously, through which neutrino mass can be constrained with low redshift probes alone; we will consider these as well. 

Our work follows a long list of papers that have forecasted constaints on neutrino mass to come from cosmological surveys 
\citep{Brinckman:2018xy, Banerjee:2016suz, Takeuchi:2013gpa, Madhavacheril:2017onh, LoVerde:2016ahu, Boyle:2017lzt}. But it is the first to explore how the combination of CMB lensing and galaxy counts can be used to exploit the effects described in the previous paragraph to evade the impact of uncertainty about the optical depth to Thomson scattering.

We will begin by introducing our forecasting assumptions, before presenting and discussing our results.


\section{\label{sec:p2}Forecasting Method and Survey Systematics}

\subsection{\label{sec:p2_1}A. Angular Power Spectra}

We use the observed galaxy density field in the $i$th tomographic redshift bin $g_i$ and the CMB lensing convergence $\kappa$ to construct the 2-point angular power spectra: $C_l^{\kappa \kappa}$, $C_l^{\kappa g_i}$, and $C_l^{g_i g_i}$. In the Limber approximation \cite{Limber:1954zz}, we model all angular power spectra
\begin{align}\label{eq:1}
C_l^{\alpha \beta} = \int \frac{dzH(z)}{\chi^2(z)} W^{\alpha}(z)W^{\beta}(z)P_{\delta_{\alpha} \delta_{\beta}} \left(k = \frac{l}{\chi(z)}, z \right),
\end{align}
where $\alpha, \beta \in (\kappa, g_1, ..., g_N)$, $H(z)$ is the Hubble parameter, $\chi(z)$ is the comoving angular-diameter distance to redshift $z$, $P(k, z)$ is the matter power spectrum at wavenumber $k$ and redshift $z$, and $N$ is the number of bins.  $\delta_{g}$ is the CDM-baryon density contrast $\delta_{cb}$, and $\delta_{\kappa}$ is the total matter density contrast $\delta_{cb \nu}$ including neutrinos. For the CMB lensing convergence, the redshift kernel $W^{\kappa}(z)$ is
\begin{align}
W^{\kappa}(z) = \frac{3}{2H(z)}\Omega_mH_0^2(1+z) \chi(z) \Bigg(\frac{\chi_* - \chi(z)}{\chi_*}\Bigg) \,,
\end{align}
where $\chi_*$ is the comoving distance to the last scattering surface, and $\Omega_m$ and $H_0$ are the matter density and the Hubble parameter today, respectively. For the $i$th bin galaxy density field $g_i$, the kernel is 
\begin{align}
W^{g_i}(z) = \frac{ b_i(z) dn_i/dz }{\int dz' (dn_i/dz')},
\end{align}
where $dn_i/dz$ is the redshift distribution of the galaxies in the $i$th bin. We assume the linear galaxy bias is given by $b_i(z) = B_i (1+z)$ within each bin, where $B_i$ is the overall bias amplitude in the $i$th bin \cite{Abell:2009aa} (we assume a fiducial value of $B_i=1$). Fig.~\ref{fig:dNdz} compares the CMB lensing kernel with the redshift distribution of two different LSST samples, as further described in Section B. We use the publicly available \texttt{CAMB} Boltzmann code to calculate the power spectrum $P_{\delta_{\alpha} \delta_{\beta}}(k,\ z)$ \cite{camb, Lewis:1999bs}.

\subsection{\label{sec:p2_2}B. LSST Specifications}

\begin{figure}[t]
\includegraphics[scale = 0.41]{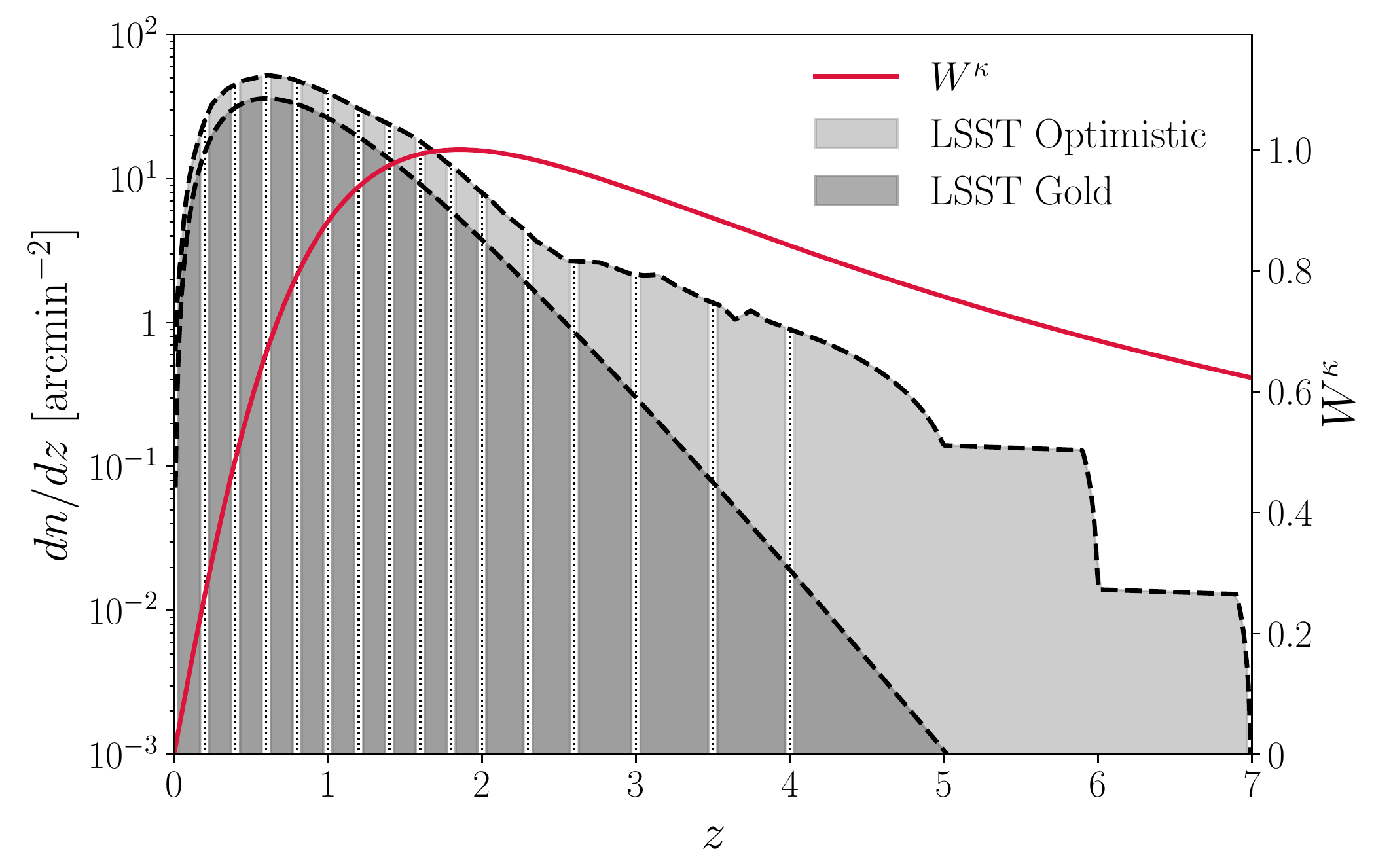}%
\caption{The redshift distribution of the CMB lensing convergence (red curve, normalized to a unit maximum) and LSST galaxy samples, both Optimistic (light gray) and Gold (dark gray). We assume 16 tomographic redshift bins in the range $0 < z < 7$, cross-correlation bin widths indicated with vertical dotted lines.}
\label{fig:dNdz}
\end{figure}

We assume two LSST number densities, as shown in Figure~\ref{fig:dNdz}. 
The first sample is the $i<25$ gold sample (henceforth referred to as ``Gold"), corresponding to $\overline{n}=40\,\mathrm{arcmin}^{-2}$ and $n(z)\propto 1/(2z_0)(z/z_0)^2 e^{-z/z_0}$ following \cite{Abell:2009aa} with $z_0 = 0.3$. As a second sample, we use a more optimistic $i<27$ magnitude cut with $S/N>5$ in the $i$ band assuming three years of observations following \cite{Gorecki:2013vv} (``Optimistic"), and add Lyman break galaxies from redshift dropouts, whose number density we estimate by scaling recent HSC observations \cite{Ono:2017wjz, Harikane:2017lcw} following \cite{Schmittfull:2017ffw}. This yields $\overline{n} \approx 66\,\mathrm{arcmin}^{-2}$ galaxies at $z=0-7$. We decompose the LSST kernel into 16 tomographic bins, with redshift edges of $z=$ [0, 0.2, 0.4, 0.6, 0.8, 1, 1.2, 1.4, 1.6, 1.8, 2, 2.3, 2.6, 3, 3.5, 4, 7], assuming that neighboring bins do not overlap. To reduce the sensitivity of our forecasts to uncertainties in non-linear modeling, including bias modeling, we keep the density perturbations in the near-linear regime by setting a $k_{\text{max}}$ limit (0.3 $h$Mpc$^{-1}$ is assumed in Figure~\ref{fig:Fig2}$-$\ref{fig:future}, with lower $k_\text{max}$ shown in Table~\ref{table1}). For each bin, we convert this to $l_{\text{max}} = k_{\text{max}}\chi(\overline{z}_i)$, where $\overline{z}_i$ is the mean redshift of the $i$th bin. Imposing $k_{\text{max}} = 0.3\ h$Mpc$^{-1}$, we find that including non-linear corrections from Halofit \cite{Smith:2002dz, Takahashi:2012em} has a negligible effect on our forecasts when all external datasets, such as primordial CMB and DESI information, are included. Hence, we use the linear matter power spectrum in all forecasts. We assume the survey area of 18,000 deg$^2$, which corresponds to $f_{\text{sky}} \approx 0.4$. Finally, we neglect any redshift space distortion effects in the LSST power spectra. 

\subsection{\label{sec:p2_3}C. CMB-S4 Specifications}

For CMB lensing, we use a CMB-S4 experiment with the following configurations: beam FWHM = 1$^{\prime}$, $\Delta_T$ = 1$\mu K^{\prime}$, and $\Delta_{E,B}$ = 1.4$\mu K^{\prime}$. We assume $f_{\text{sky}} = 0.4$, with CMB-S4 fully overlapping with the LSST \cite{Abazajian:2016yjj}. White noise is assumed, as we expect the impact of non-white noise to be small for lensing reconstruction from polarization-dominated experiments. With \texttt{quicklens} \cite{quicklens, Ade:2015zua}, we compute the minimum variance quadratic estimator lensing reconstruction on the full sky with $l_{\text{min}}^{T,E,B} = 50$, $l_{\text{max}}^{T} = 3000$, and $l_{\text{max}}^{E,B} = 5000$. We take into account the improvement from iterative lens reconstruction by rescaling the $EB$ noise \cite{Hirata:2003ka, Smith:2010gu}. In Table~\ref{table1} and~\ref{table2}, we show forecasts assuming the resulting CMB-S4 lensing reconstruction noise. For the CMB lensing convergence $\kappa$, we set $l_{\text{min}} = 30$ and $l_{\text{max}} = 2000$.

Additionally, with the CMB-S4 specifications as described above, we compute the CMB-S4 Fisher matrix, using temperature and polarization power spectra from S4, to break the parameter degeneracies. We also consider Planck primary CMB data for $l > 30$ in the region not overlapping with the CMB-S4 ($f_{\text{sky}} = 0.25$ accordingly) \cite{Abazajian:2016yjj}. Since we aim here to investigate neutrino mass constraints without $\tau$ information, no prior on the optical depth to reionization $\tau$ is included, unless we explicitly note otherwise. Here we use the unlensed CMB power spectra because the lensing auto-power spectrum $C_l^{\kappa \kappa}$ already provides nearly all the CMB lensing information \cite{Smith:2006nk} and because then the source of lensing information is entirely clear.

\subsection{\label{sec:p2_4}D. DESI Specifications}

We include the forecasted galaxy baryon acoustic oscillation (BAO) information from the Dark Energy Spectroscopic Instrument (DESI) \cite{desi} which measures the distance-redshift relation at low redshift. (We neglect RSD and other broadband sources of information in the DESI galaxy power spectrum, but assume BAO reconstruction.) Including DESI measurements significantly improves neutrino mass forecasts by better constraining $\Omega_{m}$ and further breaking parameter degeneracies. We use the expected uncertainties on the distance ratio from 18 bins in the range $0.15 < z < 1.85$ with $\Delta z = 0.1$, given in \cite{Aghamousa:2016zmz, Allison:2015qca}.

\begin{figure}[t]
\includegraphics[scale = 0.3]{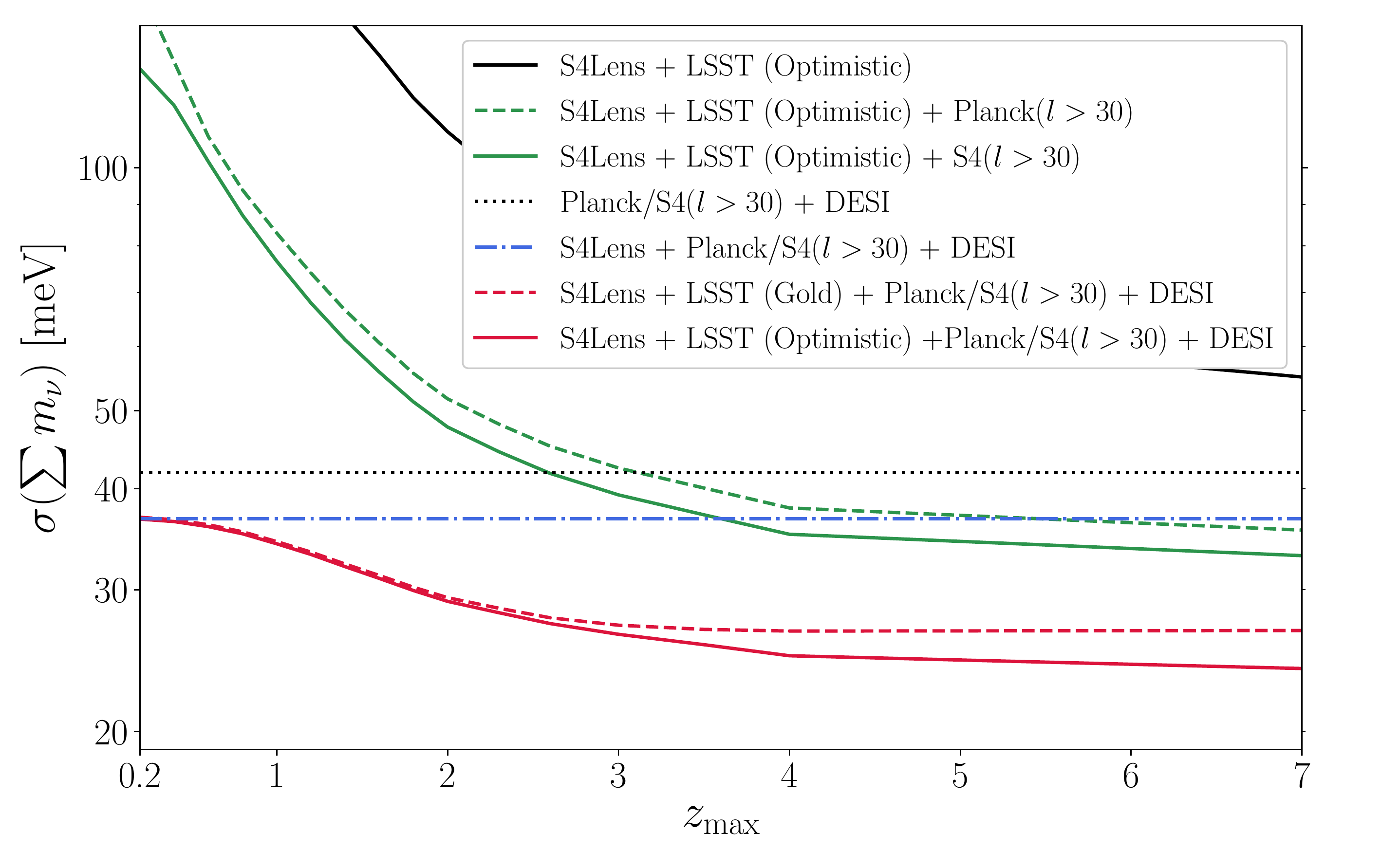}%
\caption{Forecasted 1$\sigma$ constraints on the sum of the neutrino masses without optical depth information, for different experiment configurations: CMB-S4 lensing and LSST clustering (black) + primordial CMB data (green dotted for Planck and green solid for S4) + DESI BAO measurements (red solid for LSST Optimistic and red dotted for LSST Gold). S4 primary CMB (with Planck co-added) + DESI BAO gives $\sigma(\sum{m_{\nu}})$ = 42 meV, which further tightens to 37 meV with the reconstructed CMB lensing potential included. Including the LSST galaxies at higher redshift extends the redshift lever arm and increases the volume probed, which results in a significant improvement in the constraints.}
\label{fig:Fig2}
\end{figure}

\subsection{\label{sec:p2_5}E. Fisher Matrix Analysis}

If we have $N$ tomographic galaxy redshift bins, our observables are $1+N$ (lensing-lensing and galaxy-galaxy) auto-power spectra and $N + N(N-1)/2$ (lensing-galaxy and galaxy-galaxy) cross-spectra. For the CMB lensing convergence auto-spectra, we consider the lensing reconstruction noise $N_l^{\kappa \kappa}$, and for the galaxy-galaxy auto-spectra, we take into account the shot noise $N_l^{gg} = 1/\overline{n}$. 

The Gaussian covariance matrix of the CMB lensing convergence and the LSST galaxy auto- and cross-power spectra is given by 
\begin{align}
\begin{split}
\text{Cov}^{\alpha_1 \beta_1, \alpha_2, \beta_2}_{l_a, l_b} &= \frac{\delta_{l_a, l_b}}{f_{\text{sky}}(2l_a+1)} \Big\{ (C_{l_a}^{\alpha_1 \alpha_2}+N_{l_a}^{\alpha_1 \alpha_2}) \\
&(C_{l_a}^{\beta_1 \beta_2}+N_{l_a}^{\beta_1 \beta_2}) + (C_{l_b}^{\alpha_1 \beta_2}+N_{l_b}^{\alpha_1 \beta_2}) \\
&(C_{l_b}^{\beta_1 \alpha_2}+N_{l_b}^{\beta_1 \alpha_2}) \Big\},
\end{split}
\end{align}
where $\alpha_{1,2}, \beta_{1,2} \in (\kappa, g_1, ..., g_N)$. 

We then construct the Fisher matrix
\begin{align}
F_{ij} = \sum_{\substack{\alpha_1 \beta_1, \\ \alpha_2, \nu_2}} \sum_{l} \frac{\partial C_{l}^{\alpha_1 \beta_1}}{\partial \theta_i} [\text{Cov}^{\alpha_1 \beta_1, \alpha_2, \beta_2}_{l}]^{-1} \frac{\partial C_{l}^{\alpha_2 \beta_2}}{\partial \theta_j},
\end{align}
where $\vec{\theta} = \{ B_i,\ H_0,\ \Omega_bh^2,\ \Omega_ch^2,\ n_s,\ A_s,\ \sum m_{\nu},\ \tau \}$. $B_i$ is the bias amplitude parameter of the $i$th bin. We take the fiducial values for $\tau$ and $\sum m_{\nu}$ to be 0.06 and a minimum value of 60 meV, respectively. We fix $w = -1$.

Finally, we combine the above Fisher matrix with the primordial CMB and BAO Fisher matrices and compute the marginalized constraints as 
${\rm Cov}(\theta_i, \theta_j) = (F^{-1})_{ij}$.

\setlength\extrarowheight{3pt}
\begin{table}[t]
\centering
\begin{tabular}{l | c c c}
\hline
\hline 
\phantom{.}& \multicolumn{3}{c}{\phantom{....}$\sigma(\sum{m_{\nu}})$ [meV] (Gold/Optimistic)} \phantom{....} \\
\phantom{.}$k_{\text{max}}$ \phantom{.}&\phantom{.} Lens + LSST  \phantom{.}& \phantom{.} + Planck/S4 T\&P \phantom{.}& \phantom{.}  + DESI\phantom{...}  \\
\hline
\phantom{.}0.05 \phantom{.....}
 & 307\ /\ 243 & 94\ /\ 68 & 32\ /\ 29 \\ 

\phantom{.}0.1
 & 176\ /\ 129 & 68\ /\ 53 & 31\ /\ 27 \\
 
\phantom{.}0.2
 & 107\ /\ 71 & 47\ /\ 38 & 28\ /\ 25 \\

\phantom{.}0.3
 & 84\ /\ 55 & 40\ /\ 33 & 27\ /\ 24 \\

\phantom{.}0.4 
 & 79\ /\ 49 & 38\ /\ 31 & 26\ /\ 23 \\
\hline
\end{tabular}
\caption{Forecasts of the neutrino mass constraints without optical depth information, for different LSST number densities and redshift distributions, $k_{\text{max}}$ limits, and lensing reconstruction noise levels. For different combinations of data, constraints provided on the left assume the LSST Gold sample, and those on the right assume the LSST Optimistic sample.}
\label{table1}
\end{table}

\section{\label{sec:p3}Results and Interpretation}

\begin{table}[t]
\centering
\begin{tabular}{l | c c c}
\hline
\hline 
\phantom{.}& \multicolumn{3}{c}{\phantom{....}$\sigma(\sum{m_{\nu}})$ [meV] } \\ \phantom{.}& \multicolumn{3}{c}{\phantom{.}Lens + Planck/S4 T\&P + DESI\phantom{.}} \\ \phantom{.} $k_{\text{max}}$ & \phantom{.} $\sigma(\tau) = 0.01$ & \phantom{.......} 0.005 & \phantom{...}0.002\phantom{.}\\
\hline
\phantom{...}0.3 
 \phantom{....}& \phantom{....}\ 25 \phantom{.} & \phantom{......}\ 17 & \ 12  \\ 
\hline 
\noalign{\medskip}\hline
\phantom{.}& \multicolumn{3}{c}{\phantom{....}$\sigma(\sum{m_{\nu}})$ [meV] (Gold/Optimistic)} \\ \phantom{.}& \multicolumn{3}{c}{\phantom{.}Lens + LSST + Planck/S4 T\&P + DESI\phantom{.}} \\ \phantom{.} $k_{\text{max}}$ & \phantom{.} $\sigma(\tau) = 0.01$ & \phantom{.......} 0.005 & \phantom{...}0.002\phantom{.}\\
\hline
\phantom{...}0.3 
 \phantom{....}& \phantom{....}\ 22 /\ 20 \phantom{.} & \phantom{......}\ 16 /\ 16 & \ 11 /\ 10  \\ 
\hline 
\end{tabular}
\caption{Forecasts of the neutrino mass constraints with different flat priors on the optical depth assumed. \textit{Top}: Combining CMB-S4 lensing, S4 primary CMB (with Planck co-added), and DESI BAO information. \textit{Bottom}: LSST clustering added. As in Table~\ref{table1}, numbers on the left assume the Gold sample, and those on the right assume the Optimistic sample.}
\label{table2}
\end{table}

\begin{figure}[b]
\includegraphics[scale = 0.31]{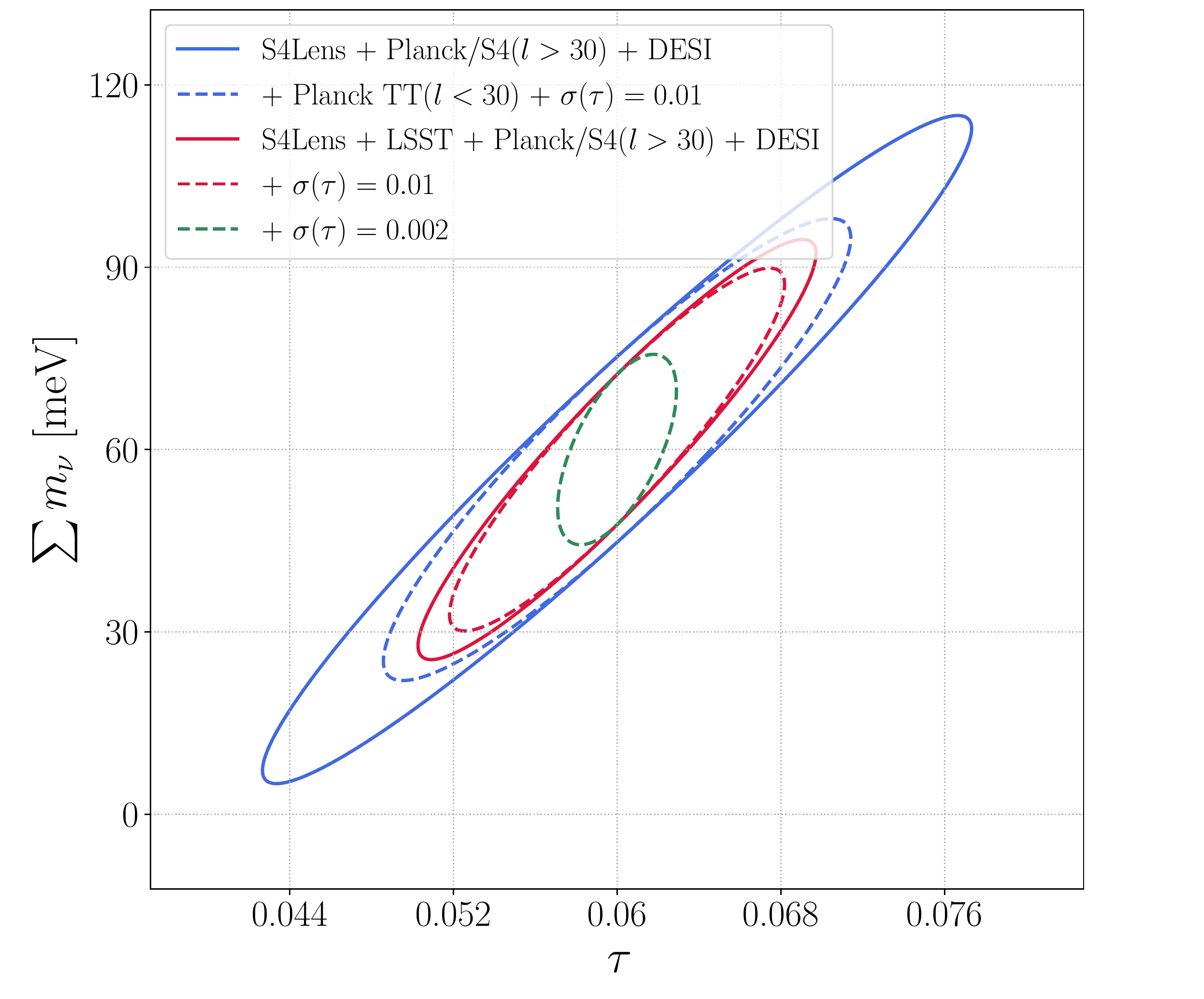}%
\caption{1$\sigma$ confidence ellipses in the $\tau-\sum{m_{\nu}}$ plane, with different combinations of datasets. The solid curves assume no prior on the optical depth, whereas the dotted curves include a flat prior on $\tau$. We find that the combination of LSST clustering and CMB-S4 lensing without any $\tau$ information (red solid) can achieve constraints competitive with or slightly better than the $\tau$-limited constraints possible with CMB-S4 (blue dotted).}
\label{fig:ellipse}
\end{figure}

With the Fisher matrix formalism described above, Fig.~\ref{fig:Fig2} presents forecasts of 1$\sigma$ constraints on the sum of neutrino masses, marginalized over $\Lambda$CDM parameters and linear galaxy biases in all redshift bins, for $k_{\text{max}} = 0.3\ h$Mpc$^{-1}$. No prior on the optical depth to reionization is included. With the LSST Optimistic sample split into 16 bins in the range $z = 0 - 7$, combining LSST clustering and CMB lensing from S4 gives $\sigma(\sum{m_{\nu}})$ = 55 meV. Adding the primordial CMB information (without any prior on $\tau$), we can achieve a constraint of 33 meV, corresponding to a $\approx$ 1.8$\sigma$ detection on the minimum value of $\sum{m_{\nu}}$ for the normal hierarchy. Using the parameter constraints from S4, we gain $\approx$ 7\% improvement in forecasts relative to the Planck primary CMB Fisher matrix. Hereafter in this analysis, we use S4 primary CMB information with Planck co-added. Finally, with the the DESI BAO measurements added, we can achieve $\sigma(\sum{m_{\nu}})$ = 24 meV, reaching a $\approx$ 2.5$\sigma$ measurement of the minimal sum of the neutrino mass, without any optical depth information. 

In Fig.~\ref{fig:Fig2}, we find that adding clustering information at higher redshift results in significantly better $\sum{m_{\nu}}$ constraints. A more pessimistic galaxy sample, LSST Gold, includes significantly less structures in high redshift and therefore yields only a minimal improvement in the constraints for $z > 3$. However, relative to the LSST Optimistic, the $\sigma(\sum{m_{\nu}})$ Gold sample constraints are not significantly worse when primary CMB and DESI information are included. We also consider the effect of having a broader redshift binning; with 6 bins in the same redshift range, $\sigma(\sum{m_{\nu}})$ degrades by $\approx$ 15\%. 

Table~\ref{table1} provides the 1$\sigma$ constraints on the neutrino mass with different $k_{\text{max}}$ limits, for both LSST Gold and Optimistic samples. Having just CMB lensing and LSST clustering, we find significant improvements as we assume a higher $k_{\text{max}}$. 
However, with all external datasets included, we find only moderate dependence on $k_\text{max}$, with a degradation of only $10-15\%$ when using $k_\text{max}=$ 0.1 $h$Mpc$^{-1}$ instead of $k_\text{max}=$ 0.3 $h$Mpc$^{-1}$.
The dependence on CMB sensitivity is similar: Improved CMB sensitivity improves constraints from CMB lensing and LSST clustering alone significantly, but only mildly when including all other probes.
We note that such modest improvements of the neutrino mass constraints with the S4 lensing reconstruction noise have been recognized previously \cite{Banerjee:2016suz}.

\begin{figure}[t]
\includegraphics[scale = 0.32]{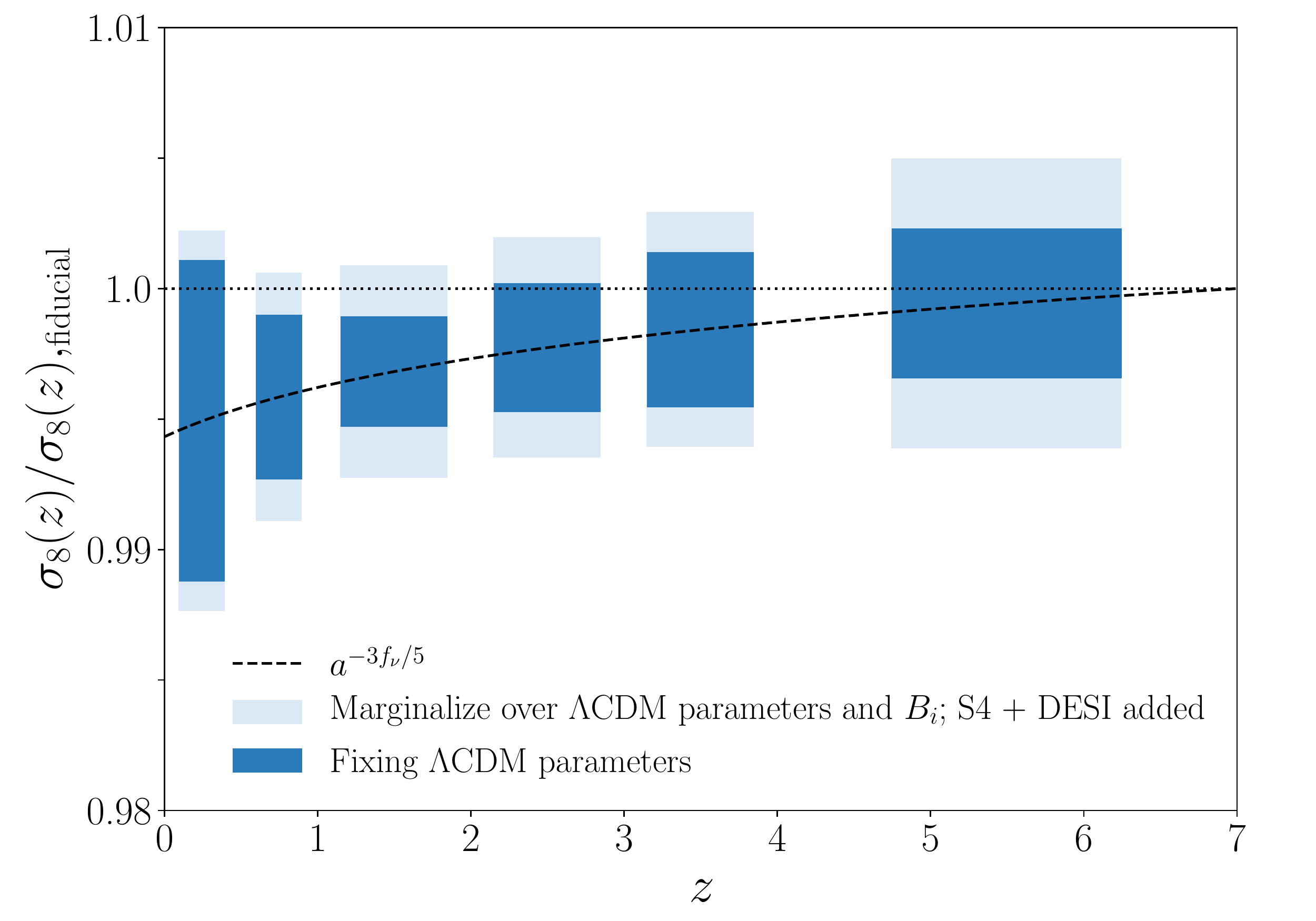}%
\caption{1$\sigma$ constraints on the matter amplitude $\sigma_8$ in 6 tomographic redshift bins, $z = 0-0.5, 0.5-1, 1-2, 2-3, 3-4, 4-7$, from the combination of LSST galaxies and CMB-S4 lensing. $k_{\text{max}}$ = 0.3 $h$Mpc$^{-1}$ is assumed. $\sigma_8$/$\sigma_{8,\text{fiducial}} = 1$ corresponds to $\sum{m_{\nu}} = 0$. Massive neutrinos suppress the growth of density fluctuations, which can be shown by how the matter density contrast scales with the scale factor: $\delta_m \propto a^{1-\frac{3}{5}f_{\nu}}$ \cite{Abazajian:2016yjj}. Assuming the minimal mass sum 60 meV, the black dotted curve plots such suppression. We either (1) marginalize over $\Lambda$CDM parameters and linear biases in each bin (light blue blocks) or (2) fix $\Lambda$CDM parameters (dark blue). In both scenarios, subpercent-level constraints on $\sigma_8$ can be achieved, leading to a significant improvement in the $\sum{m_{\nu}}$ detection.}
\label{fig:sigma8}
\end{figure}

We emphasize that the forecasts shown in Fig.~\ref{fig:Fig2} and Table~\ref{table1} assume no prior information on the optical depth. We therefore conclude that the $\tau$-less cross-correlation tomography combining LSST clustering and CMB-S4 lensing provides a different and competitive way to measure the sum of the neutrino masses.  This is better illustrated in Fig~\ref{fig:ellipse}. We obtain slightly tighter bounds on $\sum{m_{\nu}}$ and $\tau$ (red solid curve) compared to the $\tau$-limited bounds possible with CMB-S4 (blue dotted). Still, including a tight prior on $\tau$ constrains $\sum{m_{\nu}}$ better. Table~\ref{table2} summarizes the effects of the optical depth measurements on the neutrino mass constraints in our forecasts. Assuming $k_{\text{max}}$ = 0.3 $h$Mpc$^{-1}$, adding a flat prior $\sigma(\tau)$ = 0.01 improves our constraints by $15-20\%$. A better determination of $\tau$ reduces the uncertainty on the $\sum{m_{\nu}}$ detection; $\sigma(\tau)$ = 0.005 tightens our 1$\sigma$ constraint to 16 meV, and imposing the cosmic variance limit on the $\tau$ measurements brings $\sigma(\sum{m_{\nu}})$ down to 10 meV, $\approx$ 6$\sigma$ detection on the minimal sum of the neutrino masses (LSST Optimistic sample with S4 lensing noise assumed). 

What is the physical origin of these neutrino mass constraints without optical depth information? We consider two possible mechanisms by which the constraints could arise. 

\begin{figure}[t]
\includegraphics[scale = 0.33]{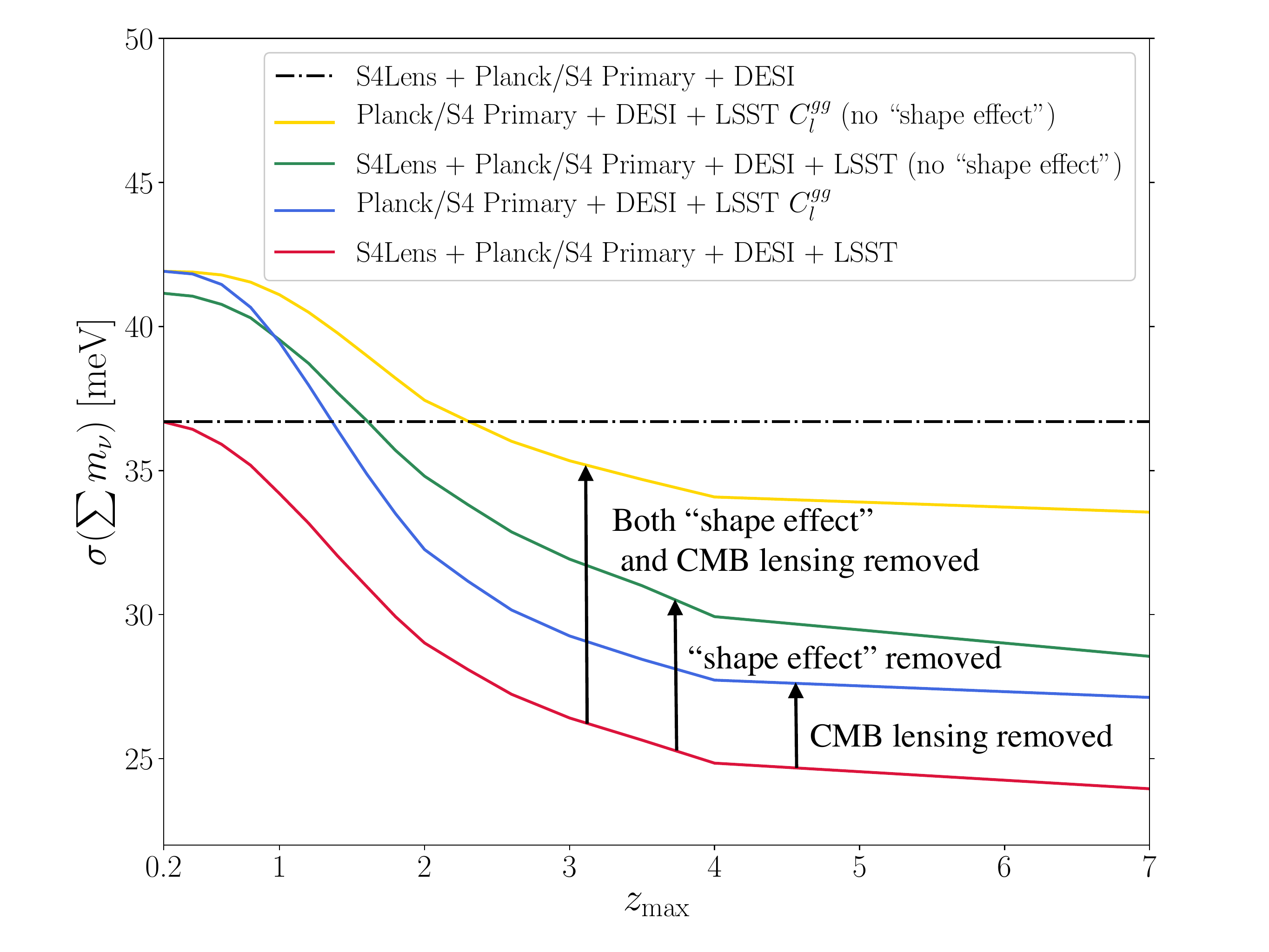}%
\caption{The relative contribution of the growth and spectrum shape effects to the $\sum{m_{\nu}}$ constraint without optical depth information. $k_{\text{max}}$ = 0.3 $h$Mpc$^{-1}$ assumed. From the full information combining both galaxies and CMB lensing (red curve), we remove either the growth effect by excluding all CMB lensing information (blue curve) or the spectrum shape effect by artificially removing the neutrino step feature (green curve). The removal of either effect substantially weakens our constraints, and removing both growth and shape effects (yellow curve) eliminates the majority of the constraining power of our data. }
\label{fig:growth}
\end{figure}

First, they could originate by probing neutrinos' effect on the growth of structure over a wider range of low redshifts. (We will henceforth refer to this as the ``growth effect''.) To illustrate this, we forecast the constraints on the amplitude of matter fluctuations $\sigma_8$ as a function of redshift, by defining a parameter $A_i$ which quantifies how the measured power spectra deviate from the standard growth of structure: $P_{mm}(k,z_i) = A_i^2 P_{mm}^{\rm fiducial}(k,z_i)$, with $A_i = 1$ for the fiducial cosmology. Following \cite{Schmittfull:2017ffw}, we consider broader redshift bins, $z = 0-0.5, 0.5-1, 1-2, 2-3, 3-4, 4-7$, and treat $A_i$ in all 6 bins as a free parameter. Marginalizing over 6 $\Lambda$CDM parameters ($ H_0,\ \Omega_bh^2,\ \Omega_ch^2,\ n_s,\ A_s,\ \tau$) and linear biases in each bin and adding external datasets, such as primary CMB and DESI, we can convert $A_i$ constraints to subpercent-level constraints on $\sigma_8$ at each redshift, as shown in Fig~\ref{fig:sigma8}. This enables us to measure (to some extent) the tiny difference between high- and low- redshift amplitudes of structure, thereby leading to a better constraint on $\sum{m_{\nu}}$. 

Since the precision to which the growth suppression alone can be measured appears moderate, we also consider other physical effects that can contribute to the constraints on neutrino mass. In particular, we consider constraints from the step-feature in the power spectrum induced by neutrino free streaming (i.e., the characteristic spectrum shape caused by growth suppression only below the free streaming scale); this should also improve with larger volume and a larger number of low-$k$ modes, as surveys extend to higher redshift. We will label this effect the ``spectrum shape effect".

In Fig.~\ref{fig:growth}, we investigate the relative contribution of the growth and spectrum shape effects to the constraints on neutrino mass without optical depth information. We begin from an analysis including the full information arising from both galaxies and CMB lensing, in which we obtain constraints shown by the red line. To understand the relative contributions, will now remove either the growth effect or the spectrum shape effect. To remove the growth effect, we simply exclude all CMB lensing information ($C_l^{\kappa \kappa}$ and $C_l^{\kappa g}$ removed); this gives the constraints shown by the blue line. To remove the spectrum shape effect, we artificially remove the neutrino step feature by matching an $\sum m_\nu=0$ power spectrum to the amplitude of the small-scale power spectrum at $k>0.1h/\mathrm{Mpc}$. This way, the whole ``featureless'' power spectrum growth is suppressed in a redshift dependent way that mimics that caused by neutrinos. This gives the constraints shown by the green line. It can be seen that in both cases, constraints are weakened substantially; the effect sizes appear comparable, though the removal of the spectrum shape effect has slightly more impact \footnote{We note that when we reduce our default $k_\mathrm{max}=0.3h/\mathrm{Mpc}$ to  $k_\mathrm{max}=0.1h/\mathrm{Mpc}$, the spectrum shape effect, which mainly arises from low $k$, becomes much more important than the growth effect, which requires 
many modes to get precise measurements of $\sigma_8(z)$.}.  Removing both the spectrum shape effect and the CMB lensing data eliminates the majority of the constraining power of our data; we thus conclude that both the shape of the galaxy power spectrum and the growth of cosmic structure, probed by high redshift galaxy and CMB lensing surveys, are responsible for the majority of our constraints on neutrino mass without optical depth information.

\begin{figure}[t]
\includegraphics[scale = 0.32]{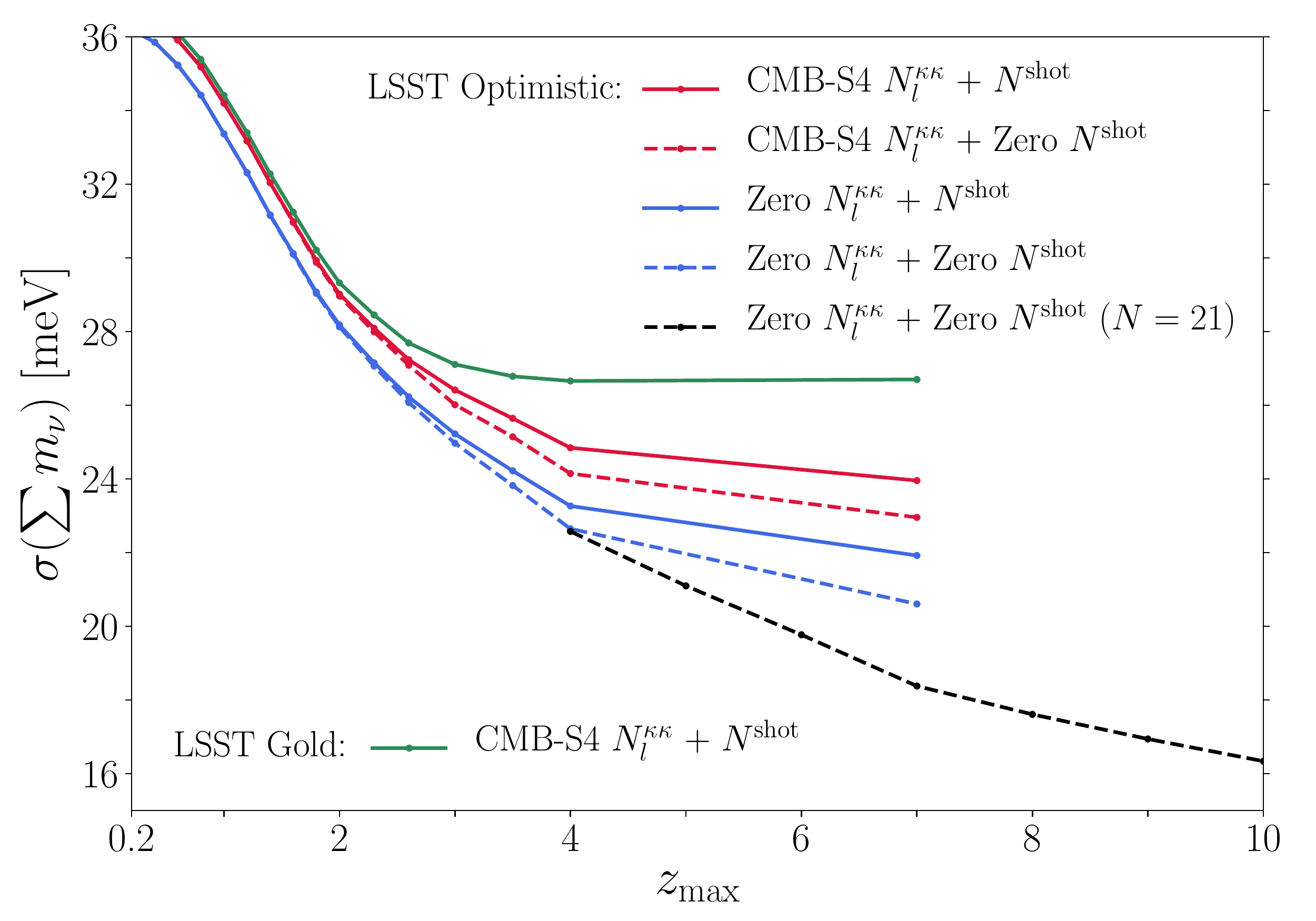}%
\caption{Forecasted 1$\sigma$ constraints on $\sum{m_{\nu}}$ with different survey configurations. The solid curves include the LSST shot noise, and the dotted curves assume zero shot noise. Having more galaxies observed in higher redshift, LSST Optimistic (red, blue, and black curves) yields tighter $\sum{m_{\nu}}$ constraints relative to LSST Gold (green). Assuming $N = 22$ in the same redshift range, we can reach up to $\sigma(\sum{m_{\nu}})$ = 16 meV. Consequently, We conclude that having more bins in high redshift tightens our constraints considerably.}
\label{fig:future}
\end{figure}

Since we find that our forecasts for neutrino mass errors do not degrade dramatically with the complete removal of CMB-S4 lensing information, we have also looked at constraints to come from a nearer-term survey with 7 times larger map noise at $\Delta_T = \Delta_{E,B}/2 = 7 \mu$K$'$. This is similar to (though not necessarily equal to) the white noise level expected from the Simons Observatory, also to be situated on the Atacama Plateau, with a survey coverage of $f_{\rm sky} \simeq 0.4$ \citep{Ade:2018sbj}. For $k_{\rm max} = 0.3h$/Mpc, our forecast for $\sigma(m_\nu)$/meV from CMB temperature, polarization and lensing combined with DESI BAO and the LSST Gold (Optimistic) sample degrades from 27 to 29 (24 to 26) when S4 is replaced with this nearer-term, noisier survey. 

Though we believe we have explained the origin of most of the combined probes' constraining power, other effects may contribute to some degree as well, such as: improved constraints on cosmological parameters such as the matter density, which may break degeneracies with neutrino mass, or constraints on the geometric factors probed by the relevant power spectra. We defer a detailed analysis attempting to quantify the impact of these other effects to future work.

The analyses described in this paper might provide the best prospects for improved constraints in future experiments, since improving optical depth constraints further may be difficult. Fig.~\ref{fig:future} explores possible improvements to our constraints, and shows that our forecasts are moderately limited by the CMB lensing reconstruction noise and the galaxy shot noise. Even though our LSST galaxy samples extend to $z = 7$, we consider one broad redshift bin for $z = 4-7$, and only modest improvements can be achieved by including this broad bin. The black dotted curve in Fig.~\ref{fig:future} assumes $N = 22$, with finer bins in high redshift: $\Delta z = 1$ in the range $z = 4-10$ (flat $dn/dz$ assumed for $z > 7$). Then, assuming zero lensing noise and shot noise, we can achieve $\sigma(\sum{m_{\nu}})$ = 16 meV from the combination of LSST galaxies and CMB-S4 lensing (all external datasets also added). This suggests that our forecasts are primarily limited by the redshift extent of the galaxy surveys. Neutrino mass constraints thus provide some motivation for extending galaxy surveys to higher redshift, though the improvements are fairly slow and the analyses will be very challenging.

\section{\label{sec:p4}Conclusions and Outlook}

We have forecast that the combination of LSST clustering and CMB-S4 lensing provides competitive neutrino mass constraints without optical depth information. Following \cite{Schmittfull:2017ffw}, we use CMB lensing -- galaxy survey cross-correlations, together with auto-power spectrum information, to cancel sample variance in part and thereby break parameter degeneracies.

For $k_{\text{max}}$ = 0.3 $h$Mpc$^{-1}$, the combination of CMB-S4 lensing with LSST galaxies, with external datasets such as Planck and S4 primordial CMB information and DESI BAO measurements included, can achieve $\sigma(\sum{m_{\nu}})$ = 24 meV, corresponding to a $\approx$ 2.5$\sigma$ detection on the minimal mass 60 meV assuming the normal hierarchy. This suggests that the $\tau$-less CMB lensing cross-correlation tomography provides an (at least partially) independent and competitive way to constrain the sum of the neutrino masses. Such improvements partially originate from sub-percent level constraints on the amplitude of structure at a number of different redshifts, which allow the measurement of the tiny difference between high- and low- redshift amplitudes of structure caused by neutrinos affecting structure growth; they also, in part, originate in constraints on the shape of the galaxy power spectrum, which benefit from the large volumes probed by high redshift surveys.

We demonstrate that including LSST galaxies at higher redshift leads to tighter constraints by extending the redshift lever arm. Comparing two LSST galaxy samples, we conclude that for a more pessimistic sample that includes less galaxies at high redshift the improvements in the constraints are only minimal for $z > 3$. We also assume zero lensing reconstruction noise and galaxy shot noise and find that the redshift lever arm and tomographic binning of the galaxy surveys (and the corresponding overlap with the lensing kernel) primarily limit our forecasts. In addition, we show that better measurements of the optical depth, if attainable and added to the analyses we describe, can improve the neutrino mass constraints further; in particular, including a cosmic-variance-limited optical depth measurement tightens $\sigma(\sum{m_{\nu}})$ to 10 meV.

We caution that for our forecasts to hold, we need to be able to model the observed power spectrum in presence of massive neutrinos to better than $\sim 1 \%$ level, which corresponds to the size of the suppression due to neutrinos in the range probed by LSST galaxies (see Fig. \ref{fig:sigma8}). For comparison, the size of the quadratic $b_2$ bias \cite{Baldauf:2016sjb,Modi:2017wds}, neglected in this analysis, can be a few percent correction to the galaxy power spectrum at $k = 0.1 h$Mpc$^{-1}$ and a $\sim 20 \%$ correction at $k = 0.3 h$Mpc$^{-1}$, depending on redshift and on the mass of the host halos. We therefore anticipate needing to model and constrain scale-dependent corrections to the $\kappa g$ and $gg$ power spectra from nonlinear bias terms to better than $\sim 10 \%$ in order to achieve the required accuracy.
Moreover, in this work we have neglected super-sample variance, and systematic errors in the analysis of high redshift galaxy clustering (such as photometric redshift uncertainties), which may limit how well we constrain the growth of structure. 
A full analysis of nonlinear biasing, photometric redshift errors and other systematic limitations is therefore well-motivated.

Nevertheless, if these systematic limitations can be controlled sufficiently well, our results show that novel high-precision neutrino mass measurements at and beyond the optical depth limit will be achievable with upcoming surveys.

\section*{Acknowledgments}
We thank Anthony Challinor,  Emmanuel Schaan, Uro\v{s} Seljak, Martin White, Michael Wilson for useful discussions.
We acknowledge the use of the \texttt{pyfisher}\footnote{https://github.com/msyriac/pyfisher} code by Mathew Madhavacheril.
B.D.S. was supported by an Isaac Newton Trust early career grant and an STFC Ernest Rutherford Fellowship. S.F. was supported by the Miller Fellowship at the University of California, Berkeley.  M.S. was supported by the Jeff Bezos Fellowship at the Institute for Advanced Study.


\end{document}